\documentclass{elsart5p}

\usepackage{graphicx}

\usepackage{amssymb}

\usepackage{lineno}

\begin{document}

\begin{frontmatter}



\title{IceTop - Cosmic Ray Physics with IceCube}
\author{Tilo Waldenmaier\corauthref{cor1}} for the IceCube Collaboration
\address[cor1]{Bartol Research Institute, DPA, University of Delaware, Newark DE, 19716, U.S.A.}
\ead{tilo@bartol.udel.edu}


\author{}

\address{}

\begin{abstract}
The IceCube experiment at South Pole consists of two detector
components - the IceTop air shower array on the surface and the
neutrino telescope at depths from 1450 to 2450 meters below. Currently, 26
IceTop stations and 22 InIce strings are deployed. With the
present size of the IceTop array, it is possible to measure cosmic rays with energies ranging from 0.5 PeV to 100 PeV. Coincident
events between the IceTop and the InIce detector provide useful
cross-checks of the detector performance and furthermore make it possible to
study the cosmic-ray composition. This paper gives an overview on the
current status of IceTop.
\end{abstract}

\begin{keyword}

IceTop \sep cosmic rays \sep extensive air shower \sep composition
\PACS 95.55.Vj \sep 95.85.Ry \sep 96.50.S- \sep 96.50.sd
\end{keyword}
\end{frontmatter}

\section{Introduction}
\begin{figure}[t]
\centering
\includegraphics[width=\linewidth]{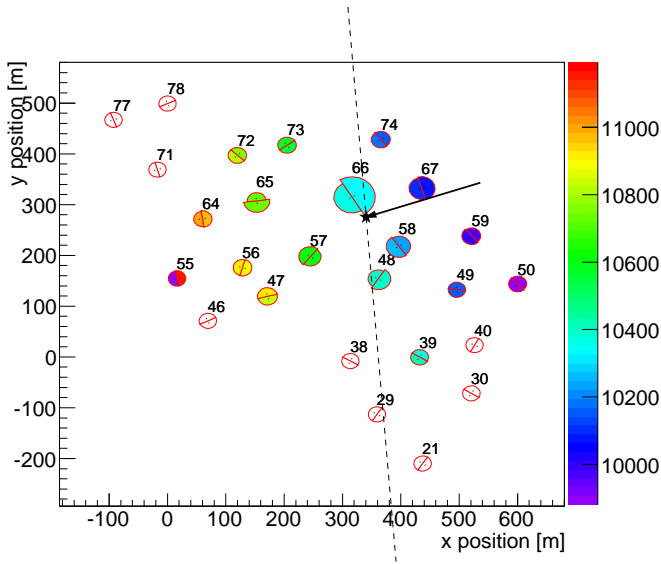}
\caption{Event display with the present layout of the 26 station
IceTop array. Each station consists of two tanks, represented by a half circle. The radius of the half circles scales with the
signal in the tanks and the different colors correspond to the
signal times. The reconstructed core
and direction are illustrated by the star and the arrow. The dashed line traces the shower plane through the core
on the surface of the array.}\label{fig:array}
\end{figure}

The IceCube neutrino telescope~\cite{Achterberg:2006} being built
at South Pole, consists of two major detector components. The actual
neutrino telescope (InIce) is buried at depths between 1450~m and
2450~m in the antarctic ice and will consist of 4800 individual photon
detectors (\textbf{D}igital \textbf{O}ptical \textbf{M}odules),
lined up on 80 strings in an hexagonal pattern. The IceTop air shower array is
located on the ice surface above the InIce detector. It consists of 80
stations, close to the position of the InIce strings, with a spacing of
125~m. Presently 26 stations are operational and are arranged as shown in
Fig.\ref{fig:array}.

The IceTop air shower array is a multi-purpose detector with several
scientific aspects and technical goals. It helps to understand the background of single muons in the InIce
detector and acts as a veto against muon bundles from cosmic ray
induced extensive air showers (EAS). Furthermore, it is used independently to
detect high energy cosmic rays with energies ranging 
from approximately 500~TeV up to about 1~EeV (in its final stage). Coincident air shower events,
measured by the IceTop and the InIce detector, make it possible to study the
cosmic-ray composition since the number of muons, which are able to
reach the InIce detector, is also sensitive to the mass of the primary
cosmic-ray particle. In addition, coincident events provide very
useful cross-checks for the precision of the direction- and energy-reconstruction in IceTop and InIce. The following
sections summarize the current status of IceTop and give some
examples for the performance of the detector.

\section{The IceTop Array}
\begin{figure}[t]
\centering
\includegraphics[width=0.9\linewidth]{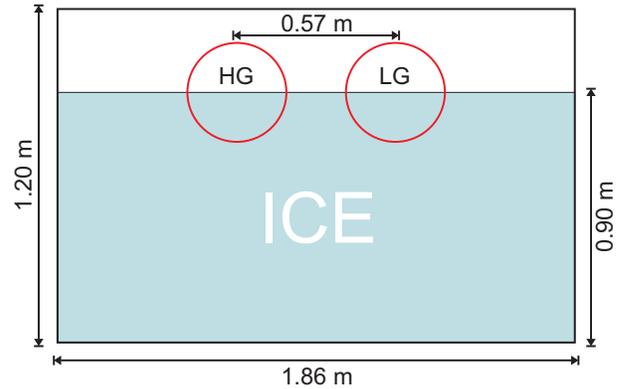}
\caption{Sketch of an IceTop tank containing a high gain (HG) and
a low gain (LG) DOM. The inner surface of the tanks is covered by
a reflective coating.}\label{fig:tank}
\end{figure}

Each of the 80 IceTop stations consists of two Ice-Cherenkov tanks
with an inner diameter of 1.86~m and an ice thickness
of 0.9~m as shown in Fig.~\ref{fig:tank}. The tanks are
buried under a thin layer of snow at a distance of about 10~m apart. The inner surface of the tanks is covered with a
diffusive coating to homogeneously reflect the induced Cherenkov
light. Two DOMs, identical to those used in the InIce detector,
are frozen half submerged in the ice surface and measure the Cherenkov
emissions in the tank. One DOM is operated at high gain (HG) and the other
at low gain (LG) in order to extend the dynamic range of the tank.

After deployment of a tank the freezing process of the ice is
controlled by a freeze control unit (FCU) which circulates and degases the
water in the tank while the temperature is slowly decreased. This way the water freezes from the
top to the bottom of the tank and the number of bubbles is significantly reduced~\cite{Stanev:2005}. The freezing process
of a tank usually takes six to eight weeks. After the freezing is
completed the FCUs are recovered and refurbished to use them in
the next deployment season.

The two tanks that make up a station are linked together by local coincidence conditions between their DOMs in order to reduce the number of accidental coincidences. A local coincidence basically occurs if two DOMs in different tanks of a station measure a signal within 250 ns. If at least six IceTop DOMs report a local coincidence within 2~$\mu$s a Simple Multiplicity Trigger (SMT) is generated, causing the data acquisition to read out the air shower event. The SMT trigger rate depends on the
size of the array and is currently of the order of 14~Hz. In
addition the IceTop array is always read out when the InIce
detector has been triggered or vice versa.

The IceTop tanks are calibrated conventionally using atmospheric muons to determine the mean signal
of a Vertical Equivalent Muon (VEM) in the tanks. This unit can be
easily compared to air shower simulations. Muon telescope measurements and
simulations show that the actual muon peak
is at about 95~\% of the peak position of the full recorded spectrum of all hits~\cite{Demiroers:2007}.

\section{Reconstruction Accuracy \& Stability}
\begin{figure}[t]
\centering
\includegraphics[width=0.7\linewidth]{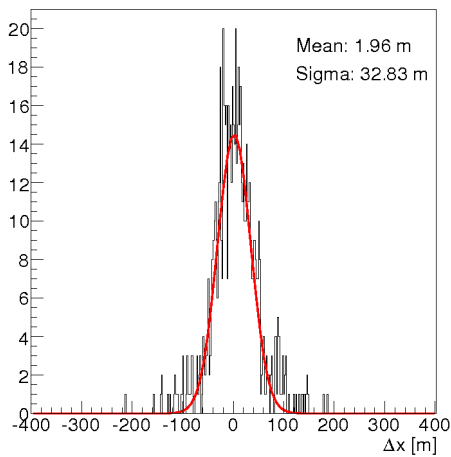}
\caption{$\Delta x$-distribution of the individually reconstructed x-coordinates of the shower cores in the two sub-arrays. See text for details.}\label{fig:deltaX}
\end{figure}

Due to the special layout of the IceTop array with two tanks (A and B)
per station, it is possible to subdivide the array into two identical
sub-arrays (A and B) with just one tank per station. This way an
air shower event can be independently reconstructed by the two
sub-arrays. The comparison of the results
gives an estimate for the stability of the reconstruction and the
uncertainties of the various shower observables. An example for the
$\Delta x$-distribution of the core reconstruction for the 16 station array in 2006 is given in
Fig.~\ref{fig:deltaX} where only contained events which fulfill the SMT trigger condition were taken
into account. The distribution in Fig.~\ref{fig:deltaX} is
fitted by a Gaussian function. Since the spread $\sigma_{\Delta x}$ of this
distribution emerges from the fluctuations of $\Delta x = x_{A} - x_{B}$ the
uncertainties $\sigma_{A,B}$ of the individually reconstructed x-coordinates
$x_{A}$ or $x_{B}$ can be obtained from
\begin{equation}
  \sigma_{\Delta x} = \sqrt{\sigma_{A}^{2} + \sigma_{B}^{2}} = \sqrt{2}\cdot\sigma_{A,B}
\end{equation}
Assuming that the actual value of $x$, using the complete array,
can be approximated as the average $\langle x\rangle=(x_{A} +
x_{B})/2$ of the results of the two sub-arrays, the actual
uncertainty of $\langle x\rangle$ becomes
\begin{equation}
  \sigma_{\langle x\rangle} = \sqrt{\left(\frac{\sigma_{A}}{2}\right)^{2}+ \left(\frac{\sigma_{B}}{2}\right)^{2}}
  = \frac{1}{2}\cdot\sigma_{\Delta x}\quad,
\end{equation}
which is half of the spread of the $\Delta x$-distribution shown
in Fig.~\ref{fig:deltaX}. The same procedure was applied to the other shower
observables revealing a core reconstruction precision of about 17~m, and an
uncertainty in the zenith and azimuth angles of
1$^{\circ}$ and 2$^{\circ}$ for the 16 station array in
2006. The reconstruction quality will increase further with the size of the
IceTop array. Since the results of this analysis may depend on the
reconstruction algorithm, the obtained uncertainties shouldn't be interpreted
as detector resolutions but are a valuable cross-check for the stability of
the reconstructions. 

\section{Single Station Coincidences}
\begin{figure}[t]
    \includegraphics[height=5.5cm,width=0.5\textwidth]{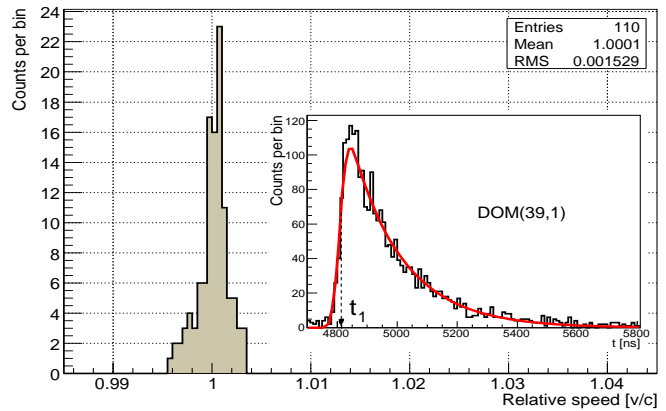}
    \caption{{\small The distribution of muon speed ($v$) relative
     to the speed of light ($c$)~\cite{Bai:2007}. The cut-in
     entry shows the time delay on one in-ice DOM and the fit.
     See text for details.}}
    \label{fig2}
\end{figure}

A precise time synchronization among all DOMs and a good knowledge about their
position is a critical requirement for any physics
analysis. Calibration with flashers and survey by hole logging
during deployment shows that the time synchronization of an InIce
string is at the level of 3~ns while the depth of the individual
DOMs are known with an accuracy of about
50~cm~\cite{Achterberg:2006}. The timing between the IceTop and
the InIce detector components can be checked by using vertical
muons which were tagged by triggering only a single station in the
IceTop detector. To ensure that the single station events are not
caused by tails of big air showers outside the array, only the
inner stations of the IceTop array are used together with the
InIce strings directly below them.  With the 16 IceTop stations
and 9 InIce strings in 2006, only stations 39 and 49 fulfill this
requirement. For these two strings the muon speed has been
individually calculated for each InIce DOM relative to the time
$t_{0}$ of the IceTop station at the surface according to
$v_i=d_i/(t_{i}-t_{0})$ where $d_i$ is the distance between the
station and the $i^{th}$ InIce DOM~\cite{Bai:2007}. In first order
the arrival times of the photons at the DOMs are exponentially
distributed due to the scattering in the ice. Since there are a
number of other independent uncertainties in the system the
exponential distribution is smeared out according to a Gaussian
function. Therefore the time distributions of individual DOMs have
been fitted with a Gaussian convoluted exponential function
\begin{equation}\label{eq:gaussexp}
\frac{dN}{dt} = \frac{1}{2}\frac{N}{\tau} e^{-\frac{t-t_{i}}{\tau}}
e^{\frac{\sigma^{2}}{2\tau^2}}\cdot \mathrm{erfc}\left(\frac{t_{i} - t +
\frac{\sigma^{2}}{\tau}}{\sqrt{2}\sigma}\right)\quad,
\end{equation}
where the time constant $\tau$ corresponds to the scattering
length in the ice and $\sigma$ is the effective time resolution of the
Gaussian function. The expression ``$\mathrm{erfc}$'' represents the
complementary error function~\cite{NumRecipes}. An example of such a fit is
given by the graph included in Fig.~\ref{fig2}. The fitted $t_{i}$ values for
each DOM correspond to the time origin of the pure exponential function and
thus to the time offset of the DOM. The distribution of muon speeds, in units
of the speed of light $c$, was determined using these time
offsets and is shown in Fig.\ref{fig2} where the data
of the two strings has been combined. Only the data of 110 out of available 120 DOMs
were used since the fit failed in the other cases due to insufficient statistics.
As can be seen in Fig.\ref{fig2} the muon speed in the ice is compatible with the speed of light
and the $RMS$ of 0.0015 reflects the uncertainties in the timing and the
location of the DOMs and of the true muon position on the
surface. If only the timing uncertainty is taken into account the $RMS$
translates to an upper limit on the timing precision of about 12~ns in 2.5 km
depths. Although the precision of this method is not as good as the
standard survey and calibration techniques, it is a useful and independent
cross-check that there is no significant deviation from expectation.

\section{Energy Spectrum \& Composition}
\begin{figure}[t]
\centering
\includegraphics[width=\linewidth]{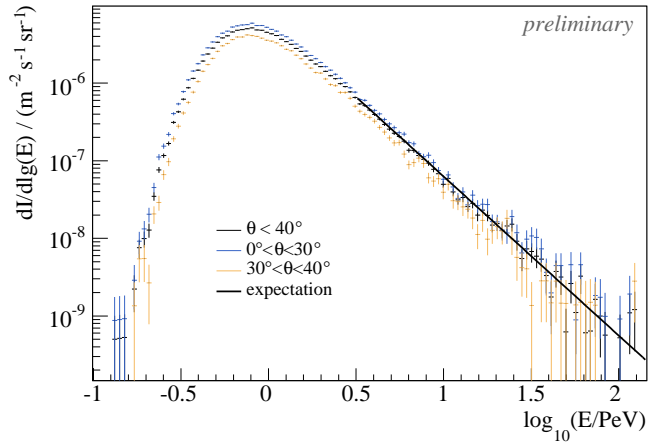}
\caption{Preliminary raw energy spectrum without acceptance correction. In the
  high energy range the spectrum already agrees well with the expected flux
  from other experiments~\cite{Klepser:2007}.}\label{fig:espectrum}
\end{figure}

The full reconstruction of an IceTop event is an iterative
process. First, the center of gravity (COG) of all tank signals is
used as a first guess for the core position. In a second step the
arrival direction of the air shower is reconstructed, using the
time information of the signals and assuming the shower front to
be a plane wave. These first guess values are taken as the seeds
for a final log-likelihood minimization which fits a lateral
distribution function to the tank signals calibrated in VEM. For
IceTop a \textbf{D}ouble \textbf{L}ogarithmic \textbf{P}arabola
(DLP) was found to describe the lateral distribution the best~\cite{Klepser:2007}. The reconstructed
signal at a distance of 100~m from the shower axis ($S_{100}$) is
used as an estimator for the primary cosmic-ray energy. The
conversion between the $S_{100}$ value and the energy is based on
a nearly linear relation between the two quantities but also
depends on the zenith angle of the shower~\cite{Klepser:2007}. For
showers with zenith angles less than $30^{\circ}$ the mean energy
for $S_{100} = 20~\mathrm{VEM}$ is approximately 10~PeV and
$S_{100} = 200~\mathrm{VEM}$ corresponds to roughly
100~PeV~\cite{Gaisser:2007}. The resulting raw energy spectrum
without any acceptance correction is shown in
Fig.~\ref{fig:espectrum}. At high energies, where the events are
detected and reconstructed with high efficiency, the slope of the
spectrum agrees well with the spectral index $\gamma = 3.05$ found
by other experiments~\cite{Klepser:2007} which is represented by
the solid line in Fig.~\ref{fig:espectrum}.

Coincident air shower events, detected by IceTop and InIce, make it possible
to determine the primary energy in three different ways, making
IceCube a hybrid cosmic-ray detector and providing a cross-check
for the different reconstruction methods and their systematic
uncertainties. Besides the method explained above, also the lateral
distribution of Cherenkov photons from the muon bundles in the InIce detector can be used to estimate the primary energy. This reconstruction is currently being developed
and provides an independent approach to estimate the primary
energy, however with a coarse energy resolution due to the primary-mass sensitivity of the muon number. Also the geometry of the
shower can be independently reconstructed by the InIce detector.
In last instance a coincident event can further be reconstructed by combining the
IceTop and the InIce data. This method leads to precise and consistent results
in the whole IceCube detector.

\begin{figure}[t]
\centering
\includegraphics[width=0.9\linewidth]{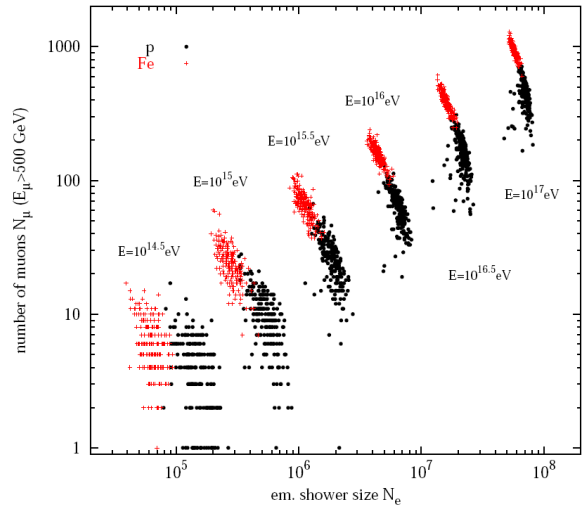}
\caption{Simulation of the correlation between the muon and the electron
  numbers in an air shower~\cite{Gaisser:2003}.}\label{fig:composition}
\end{figure}

Since the number of muons in an air shower depends on the energy
and the nature of the primary particle, the cosmic-ray composition
can be studied by correlating the muon and the electron numbers as
is suggested by the simulation study shown in
Fig.~\ref{fig:composition}. The energy increases along the diagonal of
this plot whereas the mass of the primary goes from light (proton)
to heavy (iron), roughly perpendicular to the hypothetical energy axis.
However, in real data these scatter-plots look much more fuzzy due
to the limited detector resolution and the occurrence of all kinds
of primary particles from proton to iron. The individual energy
spectra for different kinds of elemental groups can be
statistically disentangled with different techniques, for instance by a two
dimensional unfolding approach as used in~\cite{Antoni:2005}. The muon reconstruction code
for IceCube/IceTop is still under development. The cosmic-ray composition
study is one of the most important scientific goals for
IceTop.

\section{Summary}
The IceTop air shower array at South Pole grows each austral summer. After the
coming deployment season, 50~\% of the array will be
completed. In parallel to the detector construction the
reconstruction tools are being developed. An overview of the
detector performance and the existing or planned
scientific reconstruction methods has been given in this article.
The current reconstruction effort focuses on the development of a
new muon bundle reconstruction for IceCube to improve the energy
resolution for coincident air shower events and to study the
cosmic-ray composition.

\section*{Acknowledgments}
This work is supported by the U.S. National Science Foundation. Grants No. OPP-0236449 and OPP-0602679.

\bibliographystyle{elsart-num}
\bibliography{ricap07_icetop}

\begin{thebibliography}{1}
\expandafter\ifx\csname url\endcsname\relax
  \def\url#1{\texttt{#1}}\fi
\expandafter\ifx\csname urlprefix\endcsname\relax\def\urlprefix{URL }\fi

\bibitem{Achterberg:2006}
A.~Achterberg, et~al., {First year performance of the IceCube neutrino
  telescope}, Astropart. Phys. 26 (2006) 155--173.

\bibitem{Stanev:2005}
T.~Stanev, R.~Ulrich, {IceTop status in 2004}, Nucl. Phys. Proc. Suppl. 145
  (2005) 327--330.

\bibitem{Demiroers:2007}
L.~Demir{\"o}rs, et~al., {IceTop tank response to muons}, Proc. 30th Int.
  Cosmic Ray Conf., Merida (2007).

\bibitem{Bai:2007}
X.~Bai, et~al., {IceTop/IceCube coincidences}, Proc. 30th Int. Cosmic Ray
  Conf., Merida (2007).

\bibitem{NumRecipes}
W.~H. Press, S.~A. Teukolsky, W.~T. Vetterling, B.~P. Flannery, {Numerical
  Recipes, 2nd edition}, Cambridge University Press, 1992.

\bibitem{Klepser:2007}
S.~Klepser, et~al., {Lateral Distribution of Air Shower Signals and Initial
  Energy Spectrum above 1 PeV from IceTop}, Proc. 30th Int. Cosmic Ray Conf.,
  Merida (2007).

\bibitem{Gaisser:2007}
T.~Gaisser, et~al., {Performance ofthe IceTop array}, Proc. 30th Int. Cosmic
  Ray Conf., Merida (2007).

\bibitem{Antoni:2005}
T.~Antoni, et~al., Kascade measurements of energy spectra for elemental groups
  of cosmic rays: Results and open problems, Astropart. Phys. 24 (2005) 1--25.

\bibitem{Gaisser:2003}
T.~K. Gaisser, {IceTop: The surface component of IceCube}, Proc. 28th Int.
  Cosmic Ray Conf., Tsukuba (2003).

\end{thebibliography}

\end{document}